\documentclass[sigconf,nonacm]{acmart}
\usepackage{graphicx}
\usepackage{amsmath}
\usepackage{multirow}
\usepackage{booktabs}
\usepackage{tabularx}
\newcolumntype{Y}{>{\centering\arraybackslash}X}
\usepackage{makecell}
\usepackage{array}
\usepackage{float}
\usepackage{tikz}
\usetikzlibrary{shapes.geometric, arrows, calc, positioning, shadows, fit, backgrounds}
\usepackage{threeparttable}
\usepackage{siunitx}
\usepackage{xcolor}
\usepackage{placeins}
\usepackage{etoolbox}

\usetikzlibrary{positioning, backgrounds, fit, arrows.meta, calc}

\setcopyright{none}
\acmConference[SAC '27]{The 42nd ACM/SIGAPP Symposium on Applied Computing}{April 5--9, 2027}{Gwangju, Republic of Korea}
\acmYear{2027}
\copyrightyear{2027}
\renewcommand\footnotetextcopyrightpermission[1]{}

\tikzstyle{startstop} = [rectangle, rounded corners, minimum width=2.5cm, minimum height=1cm,text centered, draw=black, fill=blue!10, font=\small\sffamily]
\tikzstyle{process} = [rectangle, minimum width=2.5cm, minimum height=1cm, text centered, draw=black, fill=orange!10, font=\small\sffamily]
\tikzstyle{decision} = [diamond, minimum width=2.5cm, minimum height=1cm, text centered, draw=black, fill=green!10, font=\scriptsize\sffamily]
\tikzstyle{arrow} = [thick,->,>=stealth]
\tikzstyle{container} = [draw, rectangle, dashed, inner sep=0.5cm, rounded corners, fill=gray!5]

\definecolor{stralAccent}{HTML}{155E91}
\colorlet{stralBackground}{stralAccent!7}

\makeatletter
\patchcmd{\@mktitle@iii}
  {\@ACM@title@width=\hsize}
  {%
    \begingroup
      \setlength{\fboxsep}{8pt}%
      \setlength{\fboxrule}{0.8pt}%
      \noindent\fcolorbox{stralAccent}{stralBackground}{%
        \begin{minipage}{\dimexpr\textwidth-2\fboxsep-2\fboxrule\relax}
          \normalfont\footnotesize\raggedright\color{black}
          {\bfseries\color{stralAccent}%
            PREPRINT REVISION NOTE --- PACT TO STRAL (OCTOBER 2026)\par}
          \vspace{0.25em}%
          \begingroup
            \setlength{\leftmargini}{1.45em}%
            \begin{itemize}
              \setlength{\topsep}{0.15em}%
              \setlength{\itemsep}{0.10em}%
              \setlength{\parsep}{0pt}%
              \setlength{\partopsep}{0pt}%
              \item \textbf{Continuity.} This is a substantially revised version of
                \emph{PACT: Reducing Alert Fatigue in Low-Prevalence SOC Streams
                with Triggered Active Learning} (arXiv:2605.22324v1), renamed
                STRAL to avoid a naming collision. The core score-triggered
                hybrid active-learning mechanism is retained.
              \item \textbf{Revised methodology.} Model selection and
                pre-stream threshold calibration have been revised, and
                guarded model updates have been added.
              \item \textbf{Expanded evaluation.} The new study covers AMiner,
                Wazuh/AIT-ADS, and NAD-CD, with event-level detection and
                false-positive turnover, acquisition-policy, classifier-transfer,
                feature, and sampling analyses.
              \item \textbf{Version comparison.} The earlier BOTSv1 and
                focal-loss experiments remain in v1. Results across the two
                versions are not directly comparable because the protocols
                and operating points differ.
            \end{itemize}%
          \endgroup
        \end{minipage}%
      }%
    \endgroup
    \par\vspace{0.7em}%
    \@ACM@title@width=\hsize
  }
  {}
  {\PackageError{STRAL}{Could not insert revision note}{Check acmart class version}}
\makeatother

\begin{document}

\title[STRAL: What Adaptation Gains, Loses, and Costs in SOCs]{%
  \texorpdfstring{STRAL: What Adaptation Gains, Loses, and Costs \\ in SOC Alert Screening}%
                 {STRAL: What Adaptation Gains, Loses, and Costs in SOC Alert Screening}%
}


\author{Samuel Ndichu}
\affiliation{%
  \institution{National Institute of Information and\\
  Communications Technology (NICT)}
  \city{Koganei}
  \state{Tokyo}
  \country{Japan}
}
\email{ndichu@nict.go.jp}

\author{Tao Ban}
\affiliation{%
  \institution{National Institute of Information and\\
  Communications Technology (NICT)}
  \city{Koganei}
  \state{Tokyo}
  \country{Japan}
}
\email{bantao@nict.go.jp}

\author{Seiichi Ozawa}
\affiliation{%
  \institution{Kobe University}
  \city{Kobe}
  \state{Hyogo}
  \country{Japan}
}
\email{ozawasei@kobe-u.ac.jp}

\author{Takeshi Takahashi}
\affiliation{%
  \institution{National Institute of Information and\\
  Communications Technology (NICT)}
  \city{Koganei}
  \state{Tokyo}
  \country{Japan}
}
\email{takeshi\_takahashi@nict.go.jp}

\renewcommand{\shortauthors}{Ndichu et al.}

\begin{abstract}
Security-alert screening must balance false-alert burden against
missed important events, yet adaptation can change which events
are detected. We investigate this trade-off using STRAL, a
score-triggered active-learning controller combining ADWIN
monitoring, selective label acquisition, and bounded model
updates, with an optional historical validation Guard.
Across four chronological alert-stream settings, we compare
adaptive screening with a fixed, non-adapting baseline (Frozen),
measuring recall, false-alert burden, detection preservation,
and label-query costs.
On two AMiner settings, STRAL reduces false positives by
21.4\% (14 to 11) and 16.7\% (6 to 5), respectively,
while preserving all Frozen detections.
On Wazuh, Guard reduces false positives by 9.8\%
(522 to 471), while losing two Frozen detections.
On NAD-CD, STRAL raises recall from 36.11\% to 40.29\%,
but loses 40.1\% of Frozen detections while gaining 3,000
others; false positives increase by 151.2\%.
Guard preserves Frozen decisions on NAD-CD but requires
3,550 label queries, compared with 2,000 for STRAL.
These findings demonstrate why adaptive SOC screening
must be evaluated through detection preservation,
false-alert burden, and feedback cost alongside
aggregate recall.
\end{abstract}


\maketitle


\section{Introduction}
\label{sec:intro}

Security operations centers (SOCs) rely on intrusion-detection and
security-monitoring tools to surface activity that may warrant
investigation~\cite{bhatt2014operational,debar2001aggregation}.
These tools can generate large volumes of alerts, while the analysts
screening them have limited review capacity. Benign triggers,
redundant reports, and low-priority events compete for attention with
alerts associated with genuine attacks~\cite{dblp2023analystperspective,yang2024true}.
Downstream screening therefore faces two risks: false positives (FPs)
consume attention, while false negatives can leave important activity
unsurfaced. Classification accuracy alone does not capture either cost. Here, a
false positive is a dataset-negative alert surfaced by the downstream
screener, not necessarily an erroneous alert generated upstream.

The false-alert burden becomes substantial when negative-class events
dominate a stream. For $N_-$ negative-class events, the number
incorrectly flagged is $\mathrm{FP}=\mathrm{FPR}\,N_-$, where FPR is
the false-positive rate. An FPR of $0.1\%$ corresponds to 1,000 false
alerts per million negative-class events. We report this normalized
burden as false positives per million negative-class events (FP/1M),
alongside positive-class recall. FP/1M is a proxy for potential review
burden, not a measurement of analyst time or fatigue. Reducing it by
suppressing alerts may also suppress important events; increasing
recall may enlarge the queue requiring review.

Changes in activity, alert-generation behavior, or attack patterns
can shift the events encountered by a deployed screener. Adaptation
using newly acquired labels is one response, but labels incur cost
and selectively sampled feedback need not represent the full stream.
Moreover, a change in unlabeled prediction scores does not establish
that the relationship between features and true labels has changed.
An update can improve some screening decisions while degrading others.
The relevant question is therefore what adaptation gains, loses, and
costs, rather than whether a single aggregate metric improves.

In particular, recall does not show whether adaptation preserves the
positive events detected by the initial classifier. On the same
evaluation events, a \emph{lost detection} is a positive event flagged
by the frozen model but missed by an adaptive model; a \emph{gained
detection} is a positive event flagged only by the adaptive model.
Net recall can improve despite substantial losses among the frozen
model's detections. These terms compare decisions under two policies
on matched events; they do not describe the withdrawal of issued SOC
alerts.

We investigate this trade-off through STRAL (Score-Triggered Refinement
with Active Learning), a controller for adapting a binary security-alert
screener. STRAL monitors unlabeled prediction scores using Adaptive
Windowing (ADWIN)~\cite{bifet2007learning} and combines label queries
near the screening threshold with queries for high-scoring events.
Acquired labels support bounded model updates; a Guard variant checks
proposed updates against previously labeled evidence before deployment.
We assess how these mechanisms affect false alerts, positive-class
recall, detection preservation, and label-acquisition cost.

We evaluate four chronological security-alert screening settings:
two AMiner configurations (nominal 1\% and 5\% positive prevalence)
and a Wazuh configuration (nominal 1\%) derived from the AIT Alert
Data Set (AIT-ADS)~\cite{landauer2024aitads}, and Network Intrusion
Detection System (NIDS) Alert Data with Concept Drift
(NAD-CD)~\cite{VAARANDI2026104663}. These nominal configuration
percentages are not the measured prevalence of their evaluation
streams. For NAD-CD, we use a 10\% row subsample whose evaluation
stream contains approximately 1.9\% positive labels. In NAD-CD, positive labels denote dataset-defined important alerts,
not necessarily independently verified attacks.
An XGBoost classifier trained before stream replay, with its screening
threshold fixed on a separate calibration partition, provides the
primary frozen baseline. We compare unguarded and guarded adaptation,
label-acquisition strategies, and an alternative classifier backbone,
accounting for both screening decisions and queried labels.

\paragraph*{Contributions.}
This paper makes three contributions:
\begin{itemize}
\item \textbf{Security-grounded evaluation of adaptation.}
We evaluate Frozen, STRAL, and Guard under chronological replay.
Model selection and threshold calibration use only pre-stream data;
evaluation reports FP/1M, positive-class recall, and realized queries.

\item \textbf{Event-level detection-preservation analysis.}
Matched-event comparisons quantify frozen-baseline detections lost and
additional positives gained, together with false positives removed and
introduced. They reveal changes hidden by net recall or FP counts.

\item \textbf{Comparative analysis of adaptation mechanisms.}
We examine update schedules, acquisition policies, guarded
updates, and a Random Forest (RF) retraining variant. We distinguish
shared-backbone analyses from contextual streaming learners
with different operating points or label budgets.
\end{itemize}

\paragraph*{Research questions.}
We ask: \textbf{RQ1}, how does score-triggered adaptation change
false-alert burden, positive-class recall, and labeling cost relative
to frozen screening; \textbf{RQ2}, which frozen-baseline detections are
lost and which additional positives are gained, including when
aggregate recall improves; \textbf{RQ3}, how do update scheduling,
label acquisition, and Guard decisions change these trade-offs; and
\textbf{RQ4}, how do the outcomes differ with an RF backbone and in
contextual comparisons with other streaming-learning approaches?

\section{Related Work}
\label{sec:related_work}

\paragraph*{Downstream alert screening and SOC triage.}
Intrusion detection systems (IDSs) primarily identify suspicious
activity in traffic or logs~\cite{milenkoski2015evaluating,buczak2015data},
whereas downstream screening decides which already-generated alerts
should reach an analyst. Studies of SOC analysts and real alert
streams show that alert volume, benign triggers, and the meaning
of an alert complicate this decision~\cite{dblp2023analystperspective,yang2024true}.
Prior systems address the problem through provenance-based triage
(NoDoze)~\cite{nodoze2023}, reinforcement-learning-based alert
prioritization (AlertPro)~\cite{wang2024alertpro}, and AI-assisted
security information and event management (SIEM) workflows and
anomaly-based filtering~\cite{ban2023siem,aminanto2020threat}.
Recent work investigates automated alert classification and
triage~\cite{turcotte2025aact}, as well as learning to defer
with human feedback~\cite{jalalvand2025l2dhf}.
These approaches highlight operational trade-offs, but use different
information sources and decision mechanisms from the binary screener
and bounded label interface studied here.

\paragraph*{Active learning for alert classification.}
Interactive expert labeling has been applied to intrusion
detection~\cite{beaugnon2017ilab}, and stream clustering has
informed network IDS alert classification~\cite{vaarandi2024stream}.
Active learning has also been studied for network IDS alert
importance classification~\cite{vaarandi2024network}.
The subsequent NAD-CD study examined drift and supervised or
semi-supervised adaptation, including active learning, on a
chronological alert dataset~\cite{VAARANDI2026104663}.
FAF-BM applies semi-supervised active learning to false-alert
filtering~\cite{du2025fafbm}.
These are direct precedents for selective labeling of security alerts.

\paragraph*{Drift-aware and imbalance-sensitive stream learning.}
ADWIN detects changes in sequential observations~\cite{bifet2007learning},
and ADWIN-U extends adaptive windowing for unlabeled drift
monitoring~\cite{assis2025adwinu}. At the traffic-classification
layer, INSOMNIA and subsequent online IDS methods address
changing distributions and model updates~\cite{andresini2021insomnia,camarda2025managing}.
Streaming active learning also faces skewed class distributions:
CALMID and related approaches adapt query selection for imbalanced,
drifting streams~\cite{liu2021calmid,liu2023micfoal}, while
{\v{Z}}liobait{\.{e}} \emph{et al.} studied variable-uncertainty
acquisition under drift~\cite{zliobaite2014active}.
Such methods provide precedents for drift monitoring and selective
feedback; STRAL applies its monitor to unlabeled screening scores.

\paragraph*{Adaptive ensembles and evaluation of update costs.}
Montiel \emph{et al.} studied adaptive XGBoost for evolving data
streams~\cite{montiel2020adaptive}; Das \emph{et al.} combined
streaming anomaly discovery, selective feedback, and tree-based
ensembles~\cite{das2024effectiveness}.
These provide close precedents for updating ensembles and coupling
streaming adaptation with label acquisition, though their learning
settings differ from downstream SOC screening.
The comparison with STRAL therefore concerns evaluation scope,
not a ranking of independent learning systems.

\begin{table*}[t]
\centering
\caption{Representative work on selective feedback and security screening.}
\label{tab:related_scope}

\footnotesize
\setlength{\tabcolsep}{3pt}
\renewcommand{\arraystretch}{1.08}

\begin{tabularx}{\textwidth}{@{}
>{\raggedright\arraybackslash}p{0.20\textwidth}
>{\raggedright\arraybackslash}p{0.17\textwidth}
>{\raggedright\arraybackslash}p{0.28\textwidth}
>{\raggedright\arraybackslash}X
@{}}
\toprule
Study & Task & Learning and feedback
& Evaluation emphasis \\
\midrule

Miller et al.~\cite{miller2016reviewer}
& Malware detection
& Budgeted reviewer queries; uncertainty
  versus high-score maliciousness selection
& Detection at low FPR, review budget,
  and temporal label validity \\

Sethi and Kantardzic~\cite{sethi2017md3}
& Streaming classification
& Unlabeled margin-density monitoring;
  labels for drift verification
& Drift detection reliability and
  labeling efficiency \\

Vaarandi and Guerra-Manzanares~
\cite{vaarandi2024network}
& NIDS alert classification
& Outlier-N sampling and active learning
& Classification performance versus
  label queries \\

Camarda et al.~\cite{camarda2025managing}
& Network traffic IDS
& Drift-aware incremental RF with
  uncertainty-based label selection
& Adaptation under limited feedback \\

Drago et al.~\cite{drago2026online}
& Network traffic IDS
& Explainable unlabeled drift detection
  with AXGBoost/ARF adaptation
& Predictive performance, labeling cost,
  and explainability \\

Vaarandi and Guerra-Manzanares~
\cite{VAARANDI2026104663}
& NIDS alert classification under drift
& Supervised and active-learning-based
  semi-supervised adaptation
& Drift handling and analyst
  labeling workload \\

Waschke et al.~\cite{waschke2026ssl}
& Multiple SOC tasks
& Five semi-supervised learning methods,
  including active learning
& Prediction, labeling, training time,
  and energy costs \\

Du et al.~\cite{du2025fafbm}
& False-alert filtering
& BERT, semi-supervised learning,
  and active learning
& Filtering performance across
  heterogeneous alerts \\

\midrule

\textbf{STRAL (this work)}
& \textbf{Downstream SOC alert screening}
& Score-triggered hybrid queries;
  bounded updates; optional Guard
& \textbf{Fixed pre-stream threshold;
  FP/1M; paired detection/FP turnover;
  realized label queries} \\

\bottomrule
\end{tabularx}
\end{table*}

A recent survey reviews machine learning in SOCs~\cite{waschke2026ml4soc}.
Table~\ref{tab:related_scope} compares work on selective
feedback, drift monitoring, and alert classification.
Its acquisition, monitoring, and update mechanisms have clear
precedents. STRAL instead evaluates downstream screening at fixed,
pre-stream-calibrated thresholds through matched-event detection
and false-positive turnover, together with realized label queries.
Differences in datasets, classifiers, and feedback schedules preclude
direct performance rankings across these studies.

\section{Operational Setting and STRAL Design}
\label{sec:methods}

STRAL separates the alert-screening decision from the feedback-driven
model-update process. For each incoming event, the deployed classifier
first produces a score and a screening decision. Its pre-update score
feeds a change monitor; an eligible alarm can then initiate selective
label acquisition and a candidate model update. The unguarded controller
deploys eligible updates directly, whereas Guard checks each candidate
against previously available labels before replacement.
Figure~\ref{fig:stral_architecture} distinguishes the screening,
score-monitoring, acquisition, and update paths. The same controller
supports XGBoost continuation and RF retraining, with different
model-update operations. An update leaves the current screening
decision unchanged: it can only affect subsequent events.

\begin{figure*}[t]
\centering
\begin{tikzpicture}[
font=\sffamily\footnotesize,
node distance=4mm and 5mm,
block/.style={draw=black!65, rounded corners=1.5pt, line width=.5pt,
  align=center, text width=27mm, minimum height=10mm, inner sep=3pt},
wide/.style={block,text width=33mm},
small/.style={block,text width=23mm},
pre/.style={block,fill=black!4},
oper/.style={block,fill=black!1},
cond/.style={block,fill=black!8},
a/.style={-{Latex[length=2mm]},line width=.65pt},
l/.style={font=\sffamily\scriptsize,fill=white,inner sep=1pt}
]
\node[pre]  (hist) at (-6.9,1.35) {Pre-stream fitting and\\calibration: $f_0$, $\theta$};
\node[oper] (model) at (-2.7,1.35) {Deployed model $f_t$\\score $s_t$, fixed $\theta$};
\node[oper] (decision) at (1.5,1.35) {STREAM decision\\$\hat y_t=\mathbf 1[s_t\geq\theta]$};
\node[oper] (queue) at (5.7,1.35) {SOC screening output\\(before feedback)};
\draw[a] (hist.east) -- (model.west);
\draw[a] (model.east) -- node[l,above=1.5pt]{Score} (decision.west);
\draw[a] (decision.east) -- (queue.west);
\node[above=7mm of model,font=\sffamily\scriptsize] (incoming) {Incoming $x_t$};
\draw[a] (incoming.south) -- (model.north);

\node[cond] (adwin) at (-2.7,-0.75) {Monitor pre-update\\scores: ADWIN ($\delta=0.002$)};
\node[cond] (query) at (1.5,-0.75) {Hybrid label\\acquisition (50/50)};
\node[oper] (oracle) at (5.7,-0.75) {Offline oracle\\queried labels only};
\draw[a] (model.south) -- node[l,right]{Unlabeled $s_t$} (adwin.north);
\draw[a] (adwin) -- node[l,above]{Eligible alarm} (query);
\draw[a] (query) -- node[l,above]{Unique queries} (oracle);

\node[cond] (gate) at (-2.7,-3.05) {Unguarded: deploy\\Guard: check $V_t$};
\node[oper] (candidate) at (1.5,-3.05) {Candidate update\\XGBoost +10 rounds\\or RF refit};
\node[oper] (pending) at (5.7,-3.05) {Pending queried labels\\(at least 32)};
\draw[a] (oracle.south) -- (pending.north);
\draw[a] (pending) -- node[l,above]{Propose} (candidate);
\draw[a] (candidate) -- node[l,above]{Candidate} (gate);
\coordinate (fbLeft) at ([xshift=-.42cm]gate.west);
\coordinate (fbEnd) at ([xshift=-.68cm]model.south);
\coordinate (fbBelow) at ([yshift=-.35cm]fbEnd);
\draw[a] (gate.west) -- node[l,above]{Accept} (fbLeft) -- (fbLeft |- fbBelow)
  -- (fbBelow) -- (fbEnd);
\node[font=\sffamily\scriptsize,anchor=north,align=center] at (-2.7,-3.8)
 {Guard rejection: retain $f_t$; queries are charged};
\end{tikzpicture}
\caption{STRAL screening, feedback, and conditional model updates.}
\Description{An editable flow diagram with an upper path for pre-stream
model preparation, deployment scoring, and output to the SOC queue.
A second path sends pre-update scores to ADWIN and, conditionally, to
hybrid query selection and an offline label oracle. Pending labels
feed candidate XGBoost continuation or RF retraining. A deployment
gate either accepts the candidate directly or checks a past-only
validation set under Guard; rejection retains the current model.}
\label{fig:stral_architecture}
\end{figure*}

\subsection{Sequential screening and the frozen baseline}
At stream position $t$, $x_t$ denotes an incoming security-alert feature
vector and $y_t\in\{0,1\}$ its dataset-defined class. AIT-ADS positives
are attack-related; NAD-CD positives represent important alerts, not
necessarily verified attacks. The deployed model $f_t$ computes a
positive-class score $s_t=f_t(x_t)$, followed by the binary decision
\begin{equation}
\hat y_t=\mathbf{1}[s_t\geq\theta].
\label{eq:stral_decision}
\end{equation}
The event is scored and its decision recorded \emph{before} any label
query or model update attributable to that event. During evaluation,
this decision rule is implemented identically in deployment and Guard.

For the primary XGBoost~\cite{chen2016xgboost} experiments, the
initial model $f_0$ is trained on a pre-stream fitting partition
(FIT). The model configuration is selected using pre-stream data,
with a prespecified fallback where the selection set is uninformative.
A separate calibration partition (CAL) fixes the screening threshold
$\theta$ at a target of 1,000 FP/1M. Neither this threshold nor the
initial class weighting is recalibrated using evaluation-stream labels.
The \emph{Frozen} baseline uses $f_0$ and $\theta$ throughout the
chronological evaluation partition (STREAM) without requesting labels.
The adaptive XGBoost policies start from the same model and threshold.
Feature provenance is examined in Section~\ref{sec:experimental_setup}.

\subsection{Score-triggered label acquisition}
STREAM is processed in batches: screening decisions are recorded
before acquisition or any model update resulting from that batch.
Only when the 2,000-event cooldown has expired are pre-update scores
from the deployed model fed to ADWIN~\cite{bifet2007learning}
($\delta=0.002$). ADWIN stops consuming the batch at its first alarm;
during cooldown, it receives no scores. An alarm therefore signals a
change in observed score behavior, not verified concept drift; accepted
updates can themselves alter subsequent scores.
The recent-event buffer $\mathcal{B}_t$ holds at most 5,000 events
with the scores produced when they were screened. Buffered scores are
not recomputed when labels are selected. An eligible alarm initiates
acquisition with nominal per-trigger query allowance
\begin{equation}
q_t^{\max}=\max\!\left(32,\operatorname{round}
\left(0.01|\mathcal{B}_t|\right)\right).
\label{eq:stral_queries}
\end{equation}
This allowance applies \emph{per trigger}, not once per stream; the
actual number queried is limited by the number of eligible, previously
unqueried buffered events. There is no independent stream-wide query
cap for STRAL; for XGBoost, acquisition stops once the tree cap is reached.

The prespecified 50/50 rule assigns half of the query slots to events
with scores nearest the deployed threshold, ranked by
$|s-\theta|$, and half to the highest positive-class scores.
The 50/50 split is prespecified rather than selected using
evaluation-stream outcomes. Threshold-relative uncertainty differs from
maximum binary entropy when $\theta\neq 0.5$. Duplicate selections are
removed and unfilled slots are taken from the highest-scoring eligible
remainder. An event identifier may be queried at most once.
For each selected event, the offline oracle supplies its recorded label
immediately, after its screening decision. Labels for unqueried STREAM
events are not available to the adaptive controller. Queried labels
accumulate for a possible update; a score alarm alone does not change
the deployed model.

\subsection{Model refinement and guarded deployment}
The initial XGBoost model uses a binary logistic objective and
100 trees. Class weighting, when applied, is fixed using the
pre-stream fitting data and is not recalculated from acquired STREAM
labels. The evaluated weighting configurations and pre-stream
selection procedure are specified in
Section~\ref{subsec:model_config}.

When at least 32 queried labels are pending, unguarded STRAL
continues the current booster for up to 10 training rounds, subject
to a cap of 500 trees. Each continuation trains on that pending batch
only, not the complete history of queried labels or the original FIT
partition; there is no replay buffer or full XGBoost refit. The pending
batch is cleared on an update attempt, including a Guard rejection.
The deployment threshold remains fixed. At the tree cap, no further
XGBoost label acquisition or updates are attempted.

Guard compares a candidate $f'$ with the deployed model $f$ on a
past-only labeled set $V_t$, initially containing up to 2,000 FIT
positives and 5,000 CAL negatives. Labels queried from STREAM augment
$V_t$ \emph{after} the candidate trained on them has been decided;
its size is capped at 20,000 by removing the oldest entries. Guard
rejects a candidate if $V_t$ has fewer than 20 positives or no negatives.
Otherwise it accepts only if FP/1M does not increase and recall does
not decrease relative to $f$ on the same $V_t$, allowing a $10^{-12}$
numerical tolerance in metric comparisons. Both STREAM and Guard
classify using \texttt{float32} scores and threshold with the
same $s\geq\theta$ rule.
A rejection leaves the deployed model and threshold unchanged, but
consumes the queried labels and starts the 2,000-event cooldown; the
rejected training batch is not returned to pending updates. Its labels
remain eligible for \emph{later} Guard checks. Because $V_t$ includes
FIT positives used in the model's training lineage and selectively
sampled feedback, it is not an independent future holdout. Acceptance
cannot guarantee improved decisions on subsequent events.

\subsection{RF transfer and matched-event detection preservation}
The RF experiments retain the trigger, hybrid acquisition, cooldown,
and Guard criteria, but \emph{replace} XGBoost continuation with full
RF retraining on FIT combined with all queried labeled events;
CAL is not used in these retraining steps. RF starts from its own pre-stream model and a separately
calibrated, fixed threshold.
We therefore assess RF adaptation against \emph{RF-Frozen}, never
against XGBoost-Frozen.

For a fixed classifier family, let $\hat y_t^F$ and $\hat y_t^A$
denote Frozen and adaptive decisions on the \emph{same} STREAM events.
A \emph{lost detection} has $y_t=1$, $\hat y_t^F=1$, and
$\hat y_t^A=0$; a \emph{gained detection} has $y_t=1$,
$\hat y_t^F=0$, and $\hat y_t^A=1$. Writing their counts as $L$ and
$G$ yields $\mathrm{TP}_A=\mathrm{TP}_F-L+G$. We also count
negative-class false positives removed and introduced on paired
events. Realized oracle queries are reported separately from screening false
positives as a distinct source of potential review cost.

\section{Experimental Setup}
\label{sec:experimental_setup}

\subsection{Datasets and chronological partitions}
We use the four alert-stream settings introduced in
Section~\ref{sec:intro}: two AMiner configurations and Wazuh from
AIT-ADS~\cite{landauer2024aitads}, and a 10\% row sample of
NAD-CD~\cite{VAARANDI2026104663}. The AMiner and Wazuh percentages
in the configuration names are nominal targets, not their measured
STREAM class prevalences. NAD-CD is sampled from the canonical data
stratified by UTC day and label, with proportional allocation across hours
(seed 42). Original chronological FIT/CAL/STREAM assignments are
preserved. This label-stratified sampling constructs the offline evaluation
set; only queried STREAM labels are exposed to STRAL. For sensitivity,
we repeat the same 10\% sampling with seeds 43--46, fitting encoders,
models, and CAL thresholds separately for each draw. We also
reconstruct complete NAD-CD FIT, CAL, and STREAM partitions with
2,074,835, 1,206,555, and 8,453,783 events, respectively, and
repeat pre-stream fitting and calibration without subsampling.\footnote{Full-data
FIT/CAL/STREAM contain 28,251/14,036/160,771 positives, respectively.
The full-data unweighted XGBoost has 551 encoded features and deploys
$\theta=0.3023513854$ (float32). CAL flags 87 negatives (budget 1,192);
one float32 step lower would flag 4,142.}

The pre-stream FIT and CAL partitions precede the chronological
STREAM evaluation introduced in Section~\ref{sec:methods}.
Table~\ref{tab:stream_setup} reports positive/negative class counts
for these partitions.
For internal model selection, the final approximately 20\% of FIT
rows (in chronological order) form SELECT; candidate models are
trained on the earlier FIT rows (FIT-train). SELECT is \emph{not} an
additional independent evaluation partition. In setting order,
SELECT contains 6/567, 26/568, 57/10,101, and 786/40,710
positive/negative events. Following selection, the chosen
configuration is refitted on the full FIT partition.

\begin{table}[t]
  \centering
  \caption{Chronological class counts (positive/negative).}
  \label{tab:stream_setup}
  \footnotesize
  \setlength{\tabcolsep}{2.7pt}
  \begin{tabular}{lrrr}
    \toprule
    Setting & FIT & CAL & STREAM \\
    \midrule
    AMiner 1\% & 26/2,835 & 0/1,588 & 60/4,078 \\
    AMiner 5\% & 134/2,835 & 0/1,588 & 314/4,078 \\
    Wazuh 1\% & 100/50,687 & 0/50,000 & 5,721/481,435 \\
    NAD-CD & 2,826/204,652 & 1,400/119,250 & 16,077/829,303 \\
    \bottomrule
  \end{tabular}
\end{table}

The measured positive prevalence changes across these partitions:
from FIT to STREAM it is approximately 0.91\% to 1.45\% for
AMiner 1\%, 4.51\% to 7.15\% for AMiner 5\%, 0.20\% to
1.17\% for Wazuh, and 1.36\% to 1.90\% for NAD-CD.
These differences characterize the evaluation setup, but
cannot alone establish concept drift: partition construction,
sampling, and class composition also affect the observed rates.
The three AIT-ADS CAL partitions are constructed pre-attack-tail
benign holdouts, not naturally observed zero-attack intervals.

\begin{figure*}[t]
  \centering
  \includegraphics[width=0.9\textwidth]{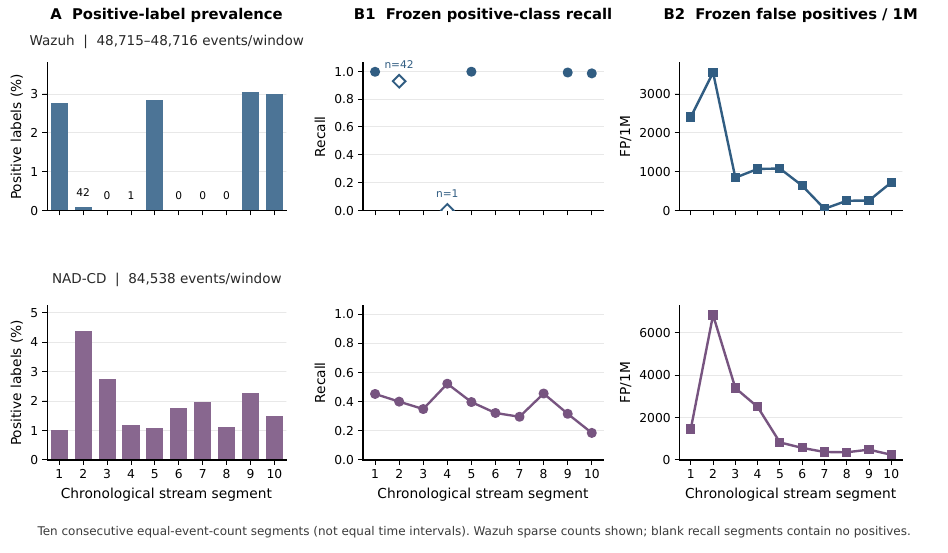}
  \caption{Chronological STREAM composition and Frozen outcomes.}
  \label{fig:stream_characterization}
  \Description{Two rows, for Wazuh and NAD-CD, each show ten consecutive
  chronological stream segments containing approximately equal numbers
  of events. Columns show positive-label prevalence,
  Frozen positive-class recall, and false
  positives per million negative-class events. Wazuh has no positives
  in four segments, so recall is undefined there; segments with one or
  42 positives have hollow markers and sparse-count annotations. NAD-CD shows substantial variation
  in prevalence and frozen screening performance. The horizontal axis
  denotes replay order rather than elapsed time.}
\end{figure*}

Figure~\ref{fig:stream_characterization} divides each evaluation
stream into ten consecutive segments containing approximately equal
numbers of events. They follow chronological replay order, not equal
intervals of elapsed time. Wazuh has 5,678 of its 5,721 positive-class
events in segments 1, 5, 9, and 10; segments 3 and 6--8 contain no
positives, so their recall is undefined. Recall in segment 4 is estimated
from a single positive event, while segment 2 contains only 42.
By contrast, NAD-CD has positive events in all ten segments; their
prevalence ranges from 1.00\% to 4.39\%, and Frozen recall ranges from
0.185 to 0.522. The Frozen false-positive burden also varies. These plots characterize temporal
variation in labels and screening outcomes; they do not, on their own,
verify concept drift or explain the changes.

\subsection{Feature provenance and leakage risks}
The settings use different operational alert features. The AMiner
streams represent host-log detector alerts using eight logical fields:
detector type, persistence filename, detector message, log resource,
affected-path root, counts of affected paths and values, and detector
training mode. These inputs produce 89 encoded model features in each
setting. Attack signatures, attack-window markers, and host identifiers
are not model inputs. Categorical encodings are fitted on pre-stream
training examples and applied without refitting during STREAM.

NAD-CD uses five alert attributes: alert count, protocol, external
and internal ports, and \texttt{SignatureID}. Fitting categorical
encoding on FIT produces 380 model features. The target \texttt{Label},
raw IP addresses, timestamps, and signature text are excluded.
\texttt{SignatureID} poses a shortcut-learning risk: in the
sampled dataset, each of the 781 observed identifiers is associated
with only one recorded label class. However, FIT contains only 361
distinct identifiers; 392 of the 695 identifiers in STREAM are absent
from FIT and map to
all-zero signature indicators under one-hot encoding. Such unseen
identifiers occur in 7,658 of 16,077 STREAM positives, so sample-wide
label purity does not imply that all positives can be recognized from
FIT-learned signature indicators. 
To examine sensitivity to upstream identifiers, we
performed a post hoc NAD-CD feature-exclusion analysis
without \texttt{SignatureID}. The four remaining alert
attributes yield 19 encoded features. The same sampled
FIT/CAL/STREAM events and chronology are retained,
while pre-stream model selection and calibration are
repeated. The selected XGBoost configuration is W2
rather than the operational-feature U configuration.
Thus, this analysis does not isolate the causal
contribution of \texttt{SignatureID}.

Wazuh uses five model inputs: numerically coded alert category,
severity, source and destination ports, and elapsed time since the
preceding alert. Category and severity originate from upstream alert or
detection-rule fields. Category codes were constructed from the full
tagged alert table before the Wazuh stream was extracted, rather than
fitted exclusively on FIT. The target label and raw IP addresses are
not model inputs. The category and severity fields can reflect
upstream rule behavior; their effects were not isolated in this study.
Consequently, the reported performance does not establish robustness
to changes in detection rules or transfer to other SOC environments.

\subsection{Model configurations, comparators, and measurement}
\label{subsec:model_config}
\paragraph*{Model selection and calibration.}
We compare binary-logistic XGBoost configurations: unweighted (U),
square-root class weighting (W1), and full-ratio weighting (W2).
For training data $T$ with negative-to-positive ratio
$\rho_T=N_-^T/N_+^T$, their positive-class weights are
$1$, $\sqrt{\rho_T}$, and $\rho_T$, respectively. Each candidate is
trained on FIT-train and assessed on SELECT using a threshold
calibrated from SELECT negatives. Among candidates meeting U's
SELECT recall, selection minimizes SELECT FP/1M, breaking ties by
higher recall and a prespecified configuration order. W1 is the
fallback when SELECT cannot support an informative comparison at the
1,000-FP/1M budget. This selects W1 for Wazuh, U for primary operational-feature
NAD-CD, and fallback W1 for both AMiner settings. Each independent
NAD-CD resample and the full-corpus experiment also selects U; the
NoSig analysis separately selects W2. Each selected configuration is
refitted on its own FIT partition, retaining its FIT-derived weight.

For deployment, set $K=\lfloor N_-^{\mathrm{CAL}}/1000\rfloor$,
corresponding to 1,000 FP/1M on CAL negatives. Calibration chooses
the \emph{lowest} \texttt{float32} threshold $\theta$ that flags at most
$K$ such events using the same single-precision $s\geq\theta$
comparison as in STREAM. CAL positives, where present, are not
used in threshold selection;
STREAM labels are unused for either selection or calibration.
The CAL constraint does not ensure the same FP/1M on STREAM. The
three AIT-ADS CAL partitions contain no positives; their calibration
cannot be assessed for positive-class recall.

Frozen, STRAL, and Guard share the same initial XGBoost model,
fixed threshold, and STREAM events within each setting. Internal comparisons vary the update
schedule (periodic every 10,000 events with random queries, or
ADWIN-triggered random queries) and the uncertainty fraction in
hybrid acquisition (0\%, 50\%, or 100\%). The 50/50 policy is the
prespecified STRAL reference, not selected on STREAM outcomes.
Periodic updates do not occur on the shorter AMiner streams;
these comparisons are not globally query-budget matched.

A Variable Uncertainty acquisition variant (VarUnc-XGB) is evaluated
with the same locked XGBoost backbone, threshold, and update engine, using
a stream-level label allowance matched to the realized STRAL count.
Its query timing differs from STRAL; this is a query-policy
comparison, not a full reproduction of the original streaming learner. An independent Massive Online Analysis (MOA)-style Variable
Uncertainty learner with Gaussian naive Bayes, Drift Detection Method
(DDM)-based resets, and a 10\% query budget
is reported only as a contextual comparison: its model, training
regime, labeling budget, and decision threshold differ.
The RF experiments use separately fitted and calibrated models;
STRAL-RF and Guard-RF are compared only with RF-Frozen.

We report positive-class recall, FP/1M, frozen-baseline detections
lost and gained, false positives removed and introduced, and
preservation of frozen true positives, all on matched STREAM events.
Unique oracle queries and accepted or rejected updates describe
feedback cost; they do not measure actual analyst time. STREAM labels
are used for evaluation, but only queried labels are exposed to
adaptive training. Results are descriptive and do not establish
statistical significance or generalization across SOC deployments.

\section{Experimental Results}
\label{sec:results}

\subsection{Screening outcomes and detection preservation}
Table~\ref{tab:stral_primary_results} compares Frozen, STRAL, and Guard
on the same chronological STREAM events and with the same initial
XGBoost model and fixed decision threshold within each setting.
Lost/gained detections and preservation refer to the counterfactual
Frozen decisions on matched positive events (Section~\ref{sec:methods});
$Q$ counts unique label queries.

\begin{table*}[t]
\centering
\caption{Primary XGBoost STREAM outcomes; turnover and preservation are Frozen-relative.}
\label{tab:stral_primary_results}
\small
\setlength{\tabcolsep}{6.5pt}
\begin{tabular}{llrrrrrrr}
\toprule
Setting & Policy & TP & FP & Recall (\%) & FP/1M & $Q$ & Lost/Gained & Pres. (\%) \\
\midrule
    AMiner 1\% & Frozen & 59 & 14 & 98.33 & 3,433.1 & 0 & -- & -- \\
     & STRAL & 59 & 11 & 98.33 & 2,697.4 & 72 & 0/0 & 100.00 \\
     & Guard & 59 & 11 & 98.33 & 2,697.4 & 72 & 0/0 & 100.00 \\
    \addlinespace[2pt]
    AMiner 5\% & Frozen & 300 & 6 & 95.54 & 1,471.3 & 0 & -- & -- \\
     & STRAL & 300 & 5 & 95.54 & 1,226.1 & 64 & 0/0 & 100.00 \\
     & Guard & 300 & 5 & 95.54 & 1,226.1 & 64 & 0/0 & 100.00 \\
    \addlinespace[2pt]
    Wazuh 1\% & Frozen & 5,673 & 522 & 99.16 & 1,084.3 & 0 & -- & -- \\
     & STRAL & 5,668 & 546 & 99.07 & 1,134.1 & 582 & 6/1 & 99.89 \\
     & Guard & 5,671 & 471 & 99.13 & 978.3 & 732 & 2/0 & 99.96 \\
    \addlinespace[2pt]
    NAD-CD & Frozen & 5,805 & 1,400 & 36.11 & 1,688.2 & 0 & -- & -- \\
     & STRAL & 6,477 & 3,517 & 40.29 & 4,240.9 & 2,000 & 2,328/3,000 & 59.90 \\
     & Guard & 5,805 & 1,400 & 36.11 & 1,688.2 & 3,550 & 0/0 & 100.00 \\
\bottomrule
\end{tabular}
\Description{Twelve canonical XGBoost STREAM outcomes on four settings. Absolute true-positive and false-positive counts; recall, normalized false-positive burden, oracle-label queries, detections lost and gained relative to Frozen, and fraction of Frozen true positives preserved.}
\end{table*}

On AMiner 1\% and 5\%, STRAL reduces false positives by three
(14 to 11) and one (6 to 5), respectively, without losing a Frozen
detection; these absolute reductions are small. Guard gives the same
screening counts. On Wazuh, unguarded STRAL instead adds 24 false
positives in net terms (522 to 546), losing six Frozen detections and
gaining one.

On NAD-CD, 3,477 positives are detected by both Frozen and
STRAL, 2,328 by Frozen alone, 3,000 by STRAL alone, and 7,272 by
neither. Recall rises from 36.11\% to 40.29\% despite preserving
only 59.9\% of Frozen's detections. False positives rise from
1,400 to 3,517: 805 are removed and 2,922 newly introduced.
These are counterfactual decisions on matched events, not
withdrawals of alerts already issued by a SOC.

\begin{figure*}[t]
\centering
\includegraphics[width=0.9\textwidth]{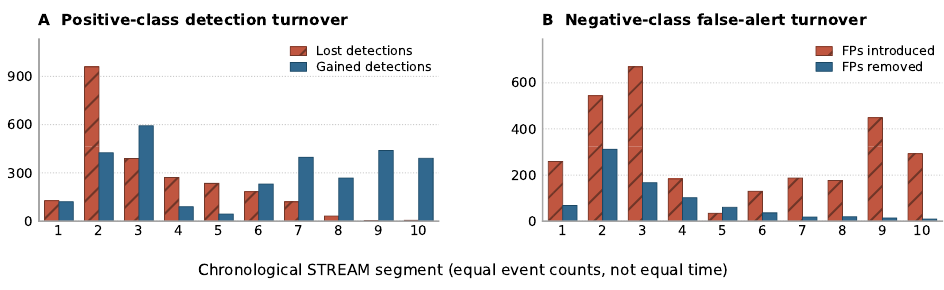}
\caption{NAD-CD positive-detection and false-positive turnover by replay segment.}
\label{fig:nad_turnover}
\Description{Two adjacent grouped-bar panels with ten chronological stream segments. Left panel: counts of Frozen-only positive detections (lost, total 2328) and STRAL-only detections (gained, total 3000). Right panel: counts of introduced false-positive decisions (total 2922) and removed false-positive decisions (total 805). The largest positive detection loss is 959 in segment 2; segment 9 has four losses and 439 gains.}
\end{figure*}

Figure~\ref{fig:nad_turnover} locates this turnover along NAD-CD replay
order. In segment 2, STRAL loses 959 Frozen detections and gains 425;
in segment 9, it loses four and gains 439. Segment 3 introduces 669
false-positive decisions and removes 167. Because segment composition
also varies (Figure~\ref{fig:stream_characterization}), these counts do not identify
a causal mechanism or independently establish concept drift.

\paragraph*{Detection turnover by signature familiarity.}
Table~\ref{tab:nad_sid_membership} stratifies the primary NAD-CD
matched-event results by whether an upstream \texttt{SignatureID}
appeared in FIT. The net gain in detections combines a regression
on FIT-seen identifiers with a larger gain on FIT-unseen identifiers.
This descriptive split does not establish that identifier familiarity
caused either change.

\begin{table}[t]
\centering
\caption{NAD-CD positive-event turnover by FIT signature membership.}
\label{tab:nad_sid_membership}
\footnotesize
\setlength{\tabcolsep}{2.5pt}
\begin{tabular*}{\columnwidth}{@{\extracolsep{\fill}}lrrr@{}}
\toprule
ID status ($n^+$) & Rec. F/S (\%) & L/G & $\Delta$TP \\
\midrule
Seen (8,419) & 52.46/47.43 & 1,698/1,274 & $-424$ \\
Unseen (7,658) & 18.12/32.44 & 630/1,726 & $+1,096$ \\
\midrule
All (16,077) & 36.11/40.29 & 2,328/3,000 & $+672$ \\
\bottomrule
\end{tabular*}
\par\vspace{2pt}{\scriptsize\emph{Note:} F/S = Frozen/STRAL recall; L/G = Frozen-relative lost/gained detections.}\par
\Description{Positive-event turnover by whether the upstream
SignatureID was observed in FIT, showing recalls, lost and gained
detections, and net true-positive changes.}
\end{table}

\subsection{Guard decisions and feedback cost}
On Wazuh, Guard accepts seven and rejects eight candidates:
false positives fall from 522 to 471, but two Frozen detections are
still lost after 732 label queries (versus 582 for STRAL). In the
primary NAD-CD setting, STRAL makes 40 accepted updates and reaches
its 500-tree cap after
2,000 queries, ending acquisition. Guard rejects all 71 candidates,
retains the initial 100-tree model and Frozen's predictions, and
continues querying, for a total of 3,550 labels. These unequal totals
arise from update and stopping behavior, not separate fixed budgets.
Guard accepts both AMiner 1\% candidates; on AMiner 5\% it accepts one
and rejects one. A safeguard thus neither guarantees future detection
preservation nor eliminates the label cost of unsuccessful updates.

\subsection{Update schedules and acquisition policies}
Table~\ref{tab:acq_comparisons} compares periodic updates,
ADWIN-triggered random queries, 0/50/100\% near-threshold acquisition,
and VarUnc-XGB. No acquisition mixture dominates on recall, FP/1M,
and preservation: for example, NAD-CD high-score-only querying raises
recall to 50.33\% but still loses 1,224 Frozen detections. On Wazuh,
Periodic applies 40 updates and queries 2,000 labels without changing
Frozen's predictions; on NAD-CD, ADWIN+random reaches 70.23\% recall
but 10,080.8 FP/1M. VarUnc-XGB shares the locked XGBoost model and
threshold, matching only STRAL's realized query count, not its query
timing or update opportunities. These are policy sensitivities rather
than equivalent-feedback end-to-end system rankings.

\begin{table*}[t]
\centering
\caption{XGBoost acquisition and update-policy sensitivity.}
\label{tab:acq_comparisons}
\begin{minipage}[t]{0.49\textwidth}
\centering
\textbf{Wazuh 1\%}\par\vspace{2pt}
{\footnotesize\setlength{\tabcolsep}{3.05pt}
\begin{tabular}{lrrrr}
\toprule
Policy & Rec. (\%) & FP/1M & $Q$ & Lost/Gained \\
\midrule
    STRAL (50/50) & 99.07 & 1,134.1 & 582 & 6/1 \\
    Periodic & 99.16 & 1,084.3 & 2,000 & 0/0 \\
    ADWIN+random & 99.18 & 4,044.2 & 932 & 0/1 \\
    High-score (0\%) & 99.07 & 980.4 & 632 & 6/1 \\
    Near-$\theta$ (100\%) & 99.13 & 1,007.4 & 632 & 2/0 \\
    VarUnc-XGB & 99.16 & 1,084.3 & 582 & 0/0 \\
\bottomrule
\end{tabular}}
\end{minipage}\hfill
\begin{minipage}[t]{0.49\textwidth}
\centering
\textbf{NAD-CD}\par\vspace{2pt}
{\footnotesize\setlength{\tabcolsep}{3.05pt}
\begin{tabular}{lrrrr}
\toprule
Policy & Rec. (\%) & FP/1M & $Q$ & Lost/Gained \\
\midrule
    STRAL (50/50) & 40.29 & 4,240.9 & 2,000 & 2,328/3,000 \\
    Periodic & 44.70 & 3,659.7 & 2,000 & 835/2,217 \\
    ADWIN+random & 70.23 & 10,080.8 & 2,000 & 1,011/6,497 \\
    High-score (0\%) & 50.33 & 4,747.4 & 2,000 & 1,224/3,511 \\
    Near-$\theta$ (100\%) & 40.54 & 3,893.6 & 2,000 & 2,333/3,045 \\
    VarUnc-XGB & 30.37 & 2,661.3 & 2,000 & 2,145/1,222 \\
\bottomrule
\end{tabular}}
\end{minipage}
\Description{Side-by-side comparison of six policies each for Wazuh and NAD-CD. Recall, false positives per million negative events, unique queried labels, and lost and gained positive detections relative to XGBoost Frozen. The published VarUncertainty acquisition policy is adapted to the same XGBoost model; it is not a native MOA end-to-end learner.}
\end{table*}

\subsection{Classifier transfer and independent context}
Table~\ref{tab:rf_cross_classifier} evaluates the separately
fitted and calibrated RF policies described in Section~\ref{sec:methods};
all turnover is relative to RF-Frozen, not XGBoost-Frozen. On the two
AMiner streams, RF adaptation leaves decisions unchanged. On Wazuh,
STRAL-RF removes some false positives while Guard-RF introduces
others. On NAD-CD, STRAL-RF markedly raises recall but increases
false positives from eight to 19,638, losing 47 RF-Frozen detections
while gaining 7,294. Guard-RF retains eight false positives yet loses
85 detections, illustrating that stable FP totals do not guarantee
positive-event preservation.

\begin{table}[t]
\centering
\caption{RF outcomes relative to independently calibrated RF-Frozen.}
\label{tab:rf_cross_classifier}
\footnotesize
\setlength{\tabcolsep}{3.2pt}
\begin{tabular}{llrrrr}
\toprule
Setting & RF policy & Rec. (\%) & FP/1M & $Q$ & Lost/Gained \\
\midrule
    Wazuh & Frozen & 99.28 & 3,387.8 & 0 & -- \\
     & STRAL-RF & 99.30 & 3,151.0 & 982 & 0/1 \\
     & Guard-RF & 99.30 & 3,535.3 & 882 & 0/1 \\
    \addlinespace[3pt]
    NAD-CD & Frozen & 29.84 & 9.6 & 0 & -- \\
     & STRAL-RF & 74.91 & 23,680.1 & 2,550 & 47/7,294 \\
     & Guard-RF & 29.68 & 9.6 & 2,150 & 85/60 \\
\bottomrule
\end{tabular}
\Description{Frozen, STRAL-RF, and Guard-RF results on Wazuh and NAD-CD using their own initial Random Forest model and calibrated threshold. Shows recall, normalized false positives, oracle queries, and matched positive detections lost and gained relative to RF Frozen.}
\end{table}

In a separate Frozen-only comparison, Isolation Forest detects
no positive events in the two AMiner settings or NAD-CD at its
calibrated operating points; low false-positive counts alone therefore
do not establish effective screening. Finally, an independent
MOA-style Variable Uncertainty learner with online Gaussian naive
Bayes and DDM attains 79.18\% recall at 14,976.4 FP/1M on NAD-CD,
querying 84,538 labels (10\% of STREAM). Its classifier, training
regime, decision rule, and feedback budget differ from STRAL; this
is contextual evidence, not a matched-budget system ranking.

\subsection{Signature-identifier exclusion sensitivity}
\label{subsec:nad_nosig}

Table~\ref{tab:nad_nosig_f32} reports a post hoc NAD-CD
analysis without \texttt{SignatureID}. On the same chronological
sample and with canonical float32 calibration, STRAL reduces the
false-positive burden but loses more Frozen detections than it gains.
Guard incurs many more label queries despite accepting only one
of 170 update proposals. Positive and negative event pairings are
computed against NoSig Frozen, not primary NAD-CD Frozen. Since the
selected initial model changes from U to W2, this is a pipeline-level
sensitivity analysis, not an isolated feature effect.

\begin{table}[!htbp]
\centering
\caption{NAD-CD without \texttt{SignatureID} (float32 calibration).}
\label{tab:nad_nosig_f32}
\footnotesize
\setlength{\tabcolsep}{2.5pt}
\begin{tabular*}{\columnwidth}{@{\extracolsep{\fill}}lrrrrr@{}}
\toprule
Policy & Rec. (\%) & FP/1M & L/G & FP R/I & $Q$ \\
\midrule
Frozen & 10.08 & 1,032.2 & -- & -- & 0 \\
STRAL & 5.25 & 282.2 & 1,031/255 & 768/146 & 1,982 \\
Guard & 11.40 & 1,006.9 & 165/378 & 367/346 & 8,482 \\
\bottomrule
\end{tabular*}
\par\vspace{2pt}{\scriptsize\emph{Note:} L/G = lost/gained positives; FP R/I = false positives removed/introduced; $Q$ = unique queries. All are NoSig Frozen-relative.}\par
\Description{Canonical float32 NoSig sensitivity analysis showing recall,
false-positive burden, matched positive and negative event turnover,
and realized oracle queries.}
\end{table}

\subsection{Sampling and full-corpus sensitivity}
\label{subsec:nad_sensitivity}
We separately replay five 10\% NAD-CD samples stratified by
day, label, and hour, fitting models and calibrating thresholds
before each stream.
Table~\ref{tab:nad_scale_sensitivity}(a) shows that STRAL raises
recall yet loses some Frozen detections and increases FP/1M in every
draw. Guard rejects all proposed updates (63--71 per draw), using
3,150--3,550 queries versus 2,000 for STRAL.

The complete 8.45-million-event STREAM uses its own FIT-trained
model and float32-calibrated threshold. In
Table~\ref{tab:nad_scale_sensitivity}(b), STRAL again gains recall
with detection losses and higher FP burden. Full-data Guard instead
accepts one of 511 proposals and gains 1,935 positives without losing
Frozen detections or changing the 712 false-positive event identities,
but requires 25,550 labels. VarUnc-XGB is included for context: its
2,000 queries match STRAL in count, not timing, and its higher recall
comes with markedly greater FP/1M.

\begin{table*}[t]
\centering
\caption{NAD-CD sampling and full-corpus sensitivity.}
\label{tab:nad_scale_sensitivity}
\setlength{\tabcolsep}{2.3pt}
\begin{minipage}[t]{0.49\textwidth}
\centering
\textbf{(a) Five 10\% sampling seeds: STRAL vs. Frozen}\par\vspace{3pt}
{\footnotesize
\begin{tabular*}{\linewidth}{@{\extracolsep{\fill}}lrrrr@{}}
\toprule
Seed & $\Delta$Rec. (pp) & $\Delta$FP/1M & L/G & Pres. (\%) \\
\midrule
42 & +4.18 & +2,553 & 2,328/3,000 & 59.9 \\
43 & +6.23 & +2,787 & 2,149/3,150 & 73.4 \\
44 & +11.15 & +2,858 & 1,329/3,122 & 78.5 \\
45 & +9.94 & +2,445 & 1,954/3,551 & 65.8 \\
46 & +16.18 & +4,015 & 1,092/3,693 & 81.0 \\
\bottomrule
\end{tabular*}}
\end{minipage}\hfill
\begin{minipage}[t]{0.49\textwidth}
\centering
\textbf{(b) Complete NAD-CD STREAM (100\%)}\par\vspace{3pt}
{\footnotesize
\begin{tabular*}{\linewidth}{@{\extracolsep{\fill}}lrrrr@{}}
\toprule
Policy & Rec. (\%) & FP/1M & L/G & $Q$ \\
\midrule
Frozen & 29.70 & 85.9 & -- & 0 \\
STRAL & 39.77 & 4,225.6 & 9,709/25,884 & 2,000 \\
Guard & 30.91 & 85.9 & 0/1,935 & 25,550 \\
VarUnc-XGB & 84.74 & 12,452.7 & 0/88,482 & 2,000 \\
\bottomrule
\end{tabular*}}
\end{minipage}
\par\vspace{3pt}
{\scriptsize\emph{Notes:} $\Delta$ values and L/G are relative to each
matched Frozen baseline; Pres. is Frozen true-positive preservation.
P1 FP/1M differences are rounded from per-seed rates. $Q$ is unique
label queries. Resampled streams overlap; VarUnc-XGB has different query timing.
STRAL queries approximately 0.24\% of sampled versus 0.024\% of full
STREAM events.\par}
\Description{Side-by-side evaluation on five separately stratified 10-percent
NAD-CD samples and on the complete 8,453,783-event evaluation stream.
The sampled results provide recall and false-positive differences, paired
positive-event turnover and preservation; the full-corpus comparison shows
recall, false-positive burden, paired positive-event turnover and label queries
for Frozen, STRAL, Guard and Variable Uncertainty XGBoost.}
\end{table*}

\section{Discussion}
\label{sec:discussion}

\subsection{Detection regressions beyond aggregate recall}
The primary NAD-CD recall gain conceals losses among positives detected
by Frozen (Table~\ref{tab:stral_primary_results}), with turnover varying
across replay segments (Figure~\ref{fig:nad_turnover}). The FIT-seen
and FIT-unseen groups also exhibit opposite net changes
(Table~\ref{tab:nad_sid_membership}). The five resamples and full
corpus preserve the qualitative pattern of net gains alongside lost
detections (Table~\ref{tab:nad_scale_sensitivity}); NoSig instead
loses recall, yet still exhibits paired turnover. These are matched
event comparisons, not causal explanations for individual errors or
evidence of withdrawn operational alerts. NAD-CD positives reflect
dataset-defined important alerts, not necessarily confirmed attacks.

\subsection{False-alert burden and the cost of safeguards}
Increased detection coverage can entail additional potential review
burden. Primary NAD-CD STRAL adds 2,117 false positives in net terms;
with independently calibrated RF, high recall likewise accompanies a
large false-positive increase (Table~\ref{tab:rf_cross_classifier}).
These within-family comparisons do not isolate a classifier-family
effect, since fitted models, thresholds, and update operations differ.
FP/1M and queries are workload proxies, not measurements of analyst
time, fatigue, or SOC capacity.

Guard illustrates the cost of selectivity rather than an assured
improvement. On the original 10\% primary NAD-CD sample it rejects
all candidates and retains Frozen decisions; on the full corpus it
accepts one candidate, retains every Frozen true positive and the same
false-positive identities, but consumes over twelve times STRAL's
queried labels. Wazuh Guard still loses two Frozen detections. Thus
past-only Guard checks provide no guarantee on future events.
Alternative acquisition mixtures and VarUnc-XGB further change the
recall--false-positive trade-off, but unequal query timing and
independent learner operating points prevent an end-to-end
budget-equivalent ranking.

\subsection{Validity and operational scope}
The study uses offline replay and dataset-provided labels, supplied
immediately and without error when queried. Analyst delay, label
disagreement, actual incident outcomes, and deployed review workloads
were not measured. NAD-CD's label-pure sampled \texttt{SignatureID}
values may permit shortcut learning; removing the feature changes the
selected model from U to W2 and reduces absolute recall, so the
ablation does not identify a feature-level causal effect. Wazuh uses
upstream rule-related fields, including categories derived from a
larger pre-extraction table; the AIT-ADS CAL sets contain no positives.
These conditions restrict transfer to other alert pipelines.

Five overlapping stratified 10\% NAD-CD draws and a complete-corpus
replay reduce dependence on one sample, but all come from the same
dataset and do not establish deployment confidence intervals or
cross-site generalization. Since the full-corpus model, encoding,
threshold, and query fraction differ, its outcomes do not isolate the
effect of dataset size. NAD-CD scores are discretized near the
threshold; all deployment calibrations use consistent float32
comparisons, yet stability at other operating points is untested.
ADWIN sensitivity, cooldown, buffer and tree-cap parameters were
not comprehensively swept.

Saved predictions support verified event-level turnover. However,
complete per-event query traces, every Guard validation-set membership,
and the exact full-data STRAL tree-cap index were not retained for
independent reconstruction. VarUnc-XGB's query-level temporal order
therefore cannot be proven solely from saved outputs. Score alarms
and prediction turnover do not establish concept drift or its causes;
Guard's historical validation set is not an independent future holdout.

\section{Conclusion}
Under fixed pre-stream screening thresholds, STRAL reduces a few
false positives without losing Frozen detections on the two AMiner
settings, but those results do not extend uniformly to Wazuh and
NAD-CD. On NAD-CD, increased recall conceals substantial paired
detection turnover and higher false-positive burden, a pattern
observed across five 10\% samples and the full corpus. Guard may
preserve or improve Frozen decisions, but at appreciable feedback
cost and without future-preservation guarantees. The NoSig and
acquisition-policy analyses reinforce that aggregate recall alone
cannot establish operational benefit. Downstream SOC evaluations
should jointly report lost and gained detections, false-alert burden,
and realized label queries.

\section*{Data Availability}
The public benchmark datasets are available from their original
providers under their terms.
Upon publication, we intend to release the experimental
implementation, preprocessing and partitioning scripts, run
configurations, saved prediction outputs, and evaluation scripts
supporting the reported results, subject to dataset licensing and
institutional requirements.

\section*{Ethical Considerations}
This study uses previously released benchmark datasets and an
offline streaming simulator. It does not involve human subjects,
live attacks, or intervention on production systems. Reported
trade-offs concern alert-screening behavior under simulated
deployment and are not used to automate real incident-response
decisions. Before deployment, operators must verify that every retained
feature is observable before the screening decision in their own
environment.

\section*{Acknowledgments}
LLMs were used for editorial purposes in this manuscript, and all outputs were inspected by the authors to ensure accuracy and originality.

\bibliographystyle{ACM-Reference-Format}
\bibliography{bibfile}

@article{bhatt2014operational,
  title={The operational role of security information and event management systems},
  author={Bhatt, Sandeep and Manadhata, Pratyusa K and Zomlot, Loai},
  journal={{IEEE} Security \& Privacy},
  volume={12},
  number={5},
  pages={35--41},
  year={2014},
  publisher={IEEE}
}

@InProceedings{debar2001aggregation,
author="Debar, Herv{\'e}
and Wespi, Andreas",
editor="Lee, Wenke
and M{\'e}, Ludovic
and Wespi, Andreas",
title="Aggregation and Correlation of Intrusion-Detection Alerts",
booktitle="Recent Advances in Intrusion Detection, LNCS volume 2212",
year="2001",
publisher="Springer Berlin Heidelberg",
address="Berlin, Heidelberg",
pages="85--103",
isbn="978-3-540-45474-8",
doi = {10.1007/3-540-45474-8_6}
}

@article{milenkoski2015evaluating,
  title={Evaluating computer intrusion detection systems: A survey of common practices},
  author={Milenkoski, Aleksandar and Vieira, Marco and Kounev, Samuel and Avritzer, Alberto and Payne, Bryan D},
  journal={ACM Computing Surveys (CSUR)},
  volume={48},
  number={1},
  articleno={12},
  numpages={41},
  year={2015},
  publisher={ACM New York, NY, USA},
  doi={10.1145/2808691},}

@inproceedings{nodoze2023,
  title={{NoDoze}: Combatting Threat Alert Fatigue with Automated Provenance Triage},
  author={Hassan, Wajih Ul and Guo, Shengjian and Li, Ding and Chen, Zhengzhang and Jee, Kangkook and Li, Zhichun and Bates, Adam},
  publisher = "The Internet Society",
  booktitle={Network and Distributed System Security Symposium (NDSS)},
  pages     = {1--15},
  numpages  = {15},
  address = {San Diego, United States},
  year={2019}
}

@inproceedings {dblp2023analystperspective,
author = {Bushra A. Alahmadi and Louise Axon and Ivan Martinovic},
title = {99\% False Positives: A Qualitative Study of {SOC} Analysts{\textquoteright} Perspectives on Security Alarms},
booktitle = {31st USENIX Security Symposium (USENIX Security 22)},
year = {2022},
isbn = {978-1-939133-31-1},
address = {Boston, MA},
pages = {2783--2800},
url = {https://www.usenix.org/conference/usenixsecurity22/presentation/alahmadi},
publisher = {USENIX Association},
month = aug
}

@inproceedings{bifet2007learning,
author = {Albert Bifet and Ricard Gavald\`a},
title = {Learning from Time-Changing Data with Adaptive Windowing},
booktitle = {Proceedings of the 2007 SIAM International Conference on Data Mining (SDM)},
year = {2007},
pages = {443-448},
publisher = {Society for Industrial and Applied Mathematics},
address   = {Philadelphia, PA, USA},
doi = {10.1137/1.9781611972771.42},
URL = {https://epubs.siam.org/doi/abs/10.1137/1.9781611972771.42},
}

@inproceedings{landauer2024aitads,
author = {Landauer, Max and Skopik, Florian and Wurzenberger, Markus},
title = {Introducing a New Alert Data Set for Multi-Step Attack Analysis},
year = {2024},
isbn = {9798400709579},
publisher = {Association for Computing Machinery},
address = {New York, NY, USA},
url = {https://doi.org/10.1145/3675741.3675748},
doi = {10.1145/3675741.3675748},
booktitle = {Proceedings of the 17th Cyber Security Experimentation and Test Workshop},
pages = {41–53},
numpages = {13},
location = {Philadelphia, PA, USA},
series = {CSET '24}
}

@inproceedings{chen2016xgboost,
author = {Chen, Tianqi and Guestrin, Carlos},
title = {XGBoost: A Scalable Tree Boosting System},
year = {2016},
isbn = {9781450342322},
publisher = {Association for Computing Machinery},
address = {New York, NY, USA},
url = {https://doi.org/10.1145/2939672.2939785},
doi = {10.1145/2939672.2939785},
booktitle = {Proceedings of the 22nd ACM SIGKDD International Conference on Knowledge Discovery and Data Mining},
pages = {785–794},
numpages = {10},
location = {San Francisco, California, USA},
series = {KDD '16}
}

@inproceedings{andresini2021insomnia,
author = {Andresini, Giuseppina and Pendlebury, Feargus and Pierazzi, Fabio and Loglisci, Corrado and Appice, Annalisa and Cavallaro, Lorenzo},
title = {INSOMNIA: Towards Concept-Drift Robustness in Network Intrusion Detection},
year = {2021},
isbn = {9781450386579},
publisher = {Association for Computing Machinery},
address = {New York, NY, USA},
url = {https://doi.org/10.1145/3474369.3486864},
doi = {10.1145/3474369.3486864},
booktitle = {Proceedings of the 14th ACM Workshop on Artificial Intelligence and Security},
pages = {111–122},
numpages = {12},
location = {Virtual Event, Republic of Korea},
series = {AISec '21}
}

@article{vaarandi2024network,
  author  = {Vaarandi, Risto and Guerra-Manzanares, Alejandro},
  title   = {Network {IDS} alert classification with active learning techniques},
  journal = {Journal of Information Security and Applications},
  volume  = {81},
  pages   = {103687},
  year    = {2024},
  doi     = {10.1016/j.jisa.2023.103687},}

@inproceedings{camarda2025managing,
  author    = {Camarda, F. and De Paola, A. and Drago, S. and Ferraro, P. and Lo Re, G.},
  title     = {Managing Concept Drift in Online Intrusion Detection Systems with Active Learning},
  booktitle = {Joint National Conference on Cybersecurity (ITASEC \& SERICS 2025)},
  series    = {CEUR Workshop Proceedings},
  publisher    = {CEUR-WS.org},
  volume    = {3962},
  pages     = {1--11},
  numpages  = {11},
  year      = {2025},
  address   = {Bologna, Italy}
}

@article{liu2023micfoal,
title = {Multiclass imbalanced and concept drift network traffic classification framework based on online active learning},
journal = {Engineering Applications of Artificial Intelligence},
volume = {117},
pages = {105607},
year = {2023},
issn = {0952-1976},
doi = {10.1016/j.engappai.2022.105607},
url = {https://www.sciencedirect.com/science/article/pii/S0952197622005978},
author = {Weike Liu and Cheng Zhu and Zhaoyun Ding and Hang Zhang and Qingbao Liu}
}

@misc{turcotte2025aact,
  author        = {Turcotte, Melissa and Labreche, Francois and Paquette, Serge-Olivier},
  title         = {Automated Alert Classification and Triage ({AACT}): An Intelligent System for the Prioritisation of Cybersecurity Alerts},
  year          = {2025},
  eprint        = {2505.09843},
  archivePrefix = {arXiv},
  primaryClass  = {cs.CR},
  url           = {https://arxiv.org/abs/2505.09843}
}

@article{wang2024alertpro,
  author  = {Wang, Xiaoyu and Yang, Xiaobo and Liang, Xueping and Zhang, Xiu and Zhang, Wei and Gong, Xiaorui},
  title   = {Combating alert fatigue with {AlertPro}: Context-aware alert prioritization using reinforcement learning for multi-step attack detection},
  journal = {Computers \& Security},
  volume  = {137},
  pages   = {103583},
  year    = {2024},
  doi     = {10.1016/j.cose.2023.103583},}

@misc{jalalvand2025l2dhf,
  author        = {Jalalvand, Fatemeh and Baruwal Chhetri, Mohan and Nepal, Surya and Paris, C{\'e}cile},
  title         = {Adaptive alert prioritisation in security operations centres via learning to defer with human feedback},
  year          = {2025},
  eprint        = {2506.18462},
  archivePrefix = {arXiv},
  primaryClass  = {cs.CR},
  url           = {https://arxiv.org/abs/2506.18462}
}

@InProceedings{beaugnon2017ilab,
author="Beaugnon, Ana{\"e}l
and Chifflier, Pierre
and Bach, Francis",
editor="Dacier, Marc
and Bailey, Michael
and Polychronakis, Michalis
and Antonakakis, Manos",
title="ILAB: An Interactive Labelling Strategy for Intrusion Detection",
booktitle="Research in Attacks, Intrusions, and Defenses",
year="2017",
publisher="Springer International Publishing",
address="Cham",
pages="120--140",
isbn="978-3-319-66332-6"
}

@article{liu2021calmid,
  author  = {Liu, Weike and Zhang, Hang and Ding, Zhaoyun and Liu, Qingbao and Zhu, Cheng},
  title   = {A comprehensive active learning method for multiclass imbalanced data streams with concept drift},
  journal = {Knowledge-Based Systems},
  volume  = {215},
  pages   = {106778},
  year    = {2021},
  doi     = {10.1016/j.knosys.2021.106778},}

@inproceedings {yang2024true,
author = {Limin Yang and Zhi Chen and Chenkai Wang and Zhenning Zhang and Sushruth Booma and Phuong Cao and Constantin Adam and Alexander Withers and Zbigniew Kalbarczyk and Ravishankar K. Iyer and Gang Wang},
title = {True Attacks, Attack Attempts, or Benign Triggers? An Empirical Measurement of Network Alerts in a Security Operations Center},
booktitle = {33rd USENIX Security Symposium (USENIX Security 24)},
year = {2024},
isbn = {978-1-939133-44-1},
address = {Philadelphia, PA},
pages = {1525--1542},
url = {https://www.usenix.org/conference/usenixsecurity24/presentation/yang-limin},
publisher = {USENIX Association},
month = aug
}

@INPROCEEDINGS{montiel2020adaptive,
  author={Montiel, Jacob and Mitchell, Rory and Frank, Eibe and Pfahringer, Bernhard and Abdessalem, Talel and Bifet, Albert},
  booktitle={2020 International Joint Conference on Neural Networks (IJCNN)}, 
  title={Adaptive XGBoost for Evolving Data Streams}, 
  year={2020},
  volume={},
  number={},
  pages={1-8},
  numpages  = {8},
  address   = {Glasgow, United Kingdom},
  publisher = {IEEE},
  doi={10.1109/IJCNN48605.2020.9207555}}

@article{ban2023siem,
  title={Breaking Alert Fatigue: {AI}-Assisted {SIEM} Framework for Effective Incident Response},
  author={Ban, Tao and Takahashi, Takeshi and Ndichu, Samuel and Inoue, Daisuke},
  journal={Applied Sciences},
  volume={13},
  number={11},
  pages={6610},
  year={2023},
  publisher={MDPI},
  doi={10.3390/app13116610},}

@inproceedings{du2025fafbm,
author = {Du, Dan and Li, Yunpeng and Cao, Yiyang and Liu, Yuling and Meng, Guozhu and Li, Ning and Han, Dongxu and Feng, Huamin},
title = {FAF-BM: An Approach for False Alerts Filtering Using BERT Model with Semi-supervised Active Learning},
year = {2025},
isbn = {978-981-96-2416-4},
publisher={Springer},
address={Singapore},
url = {https://doi.org/10.1007/978-981-96-2417-1_16},
doi = {10.1007/978-981-96-2417-1_16},
booktitle = {Science of Cyber Security: 6th International Conference, SciSec 2024, Copenhagen, Denmark, August 14–16, 2024, Proceedings},
pages = {295–312},
numpages = {18}
}

@article{assis2025adwinu,
  author  = {Assis, D. N. and Souza, V. M. A.},
  title   = {{ADWIN-U}: adaptive windowing for unsupervised drift detection on data streams},
  journal = {Knowledge and Information Systems},
  volume  = {67},
  pages   = {10005--10034},
  year    = {2025},
  doi     = {10.1007/s10115-025-02523-1},}

@article{vaarandi2024stream,
  author  = {Vaarandi, Risto and Guerra-Manzanares, Alejandro},
  title   = {Stream Clustering Guided Supervised Learning for Classifying {NIDS} Alerts},
  journal = {Future Generation Computer Systems},
  volume  = {155},
  pages   = {231--244},
  year    = {2024},
  doi     = {10.1016/j.future.2024.01.032},}

@ARTICLE{aminanto2020threat,
  author={Aminanto, Muhamad Erza and Ban, Tao and Isawa, Ryoichi and Takahashi, Takeshi and Inoue, Daisuke},
  journal={IEEE Access}, 
  title={Threat Alert Prioritization Using Isolation Forest and Stacked Auto Encoder With Day-Forward-Chaining Analysis}, 
  year={2020},
  volume={8},
  number={},
  pages={217977-217986},
  doi={10.1109/ACCESS.2020.3041837},}

@article{buczak2015data,
  title={A survey of data mining and machine learning methods for cyber security intrusion detection},
  author={Buczak, Anna L and Guven, Erhan},
  journal={IEEE Communications surveys \& tutorials},
  volume={18},
  number={2},
  pages={1153--1176},
  year={2016},
  publisher={IEEE},
  doi={10.1109/COMST.2015.2494502},}

@article{VAARANDI2026104663,
title = {NAD-CD: Addressing concept drift in network IDS alert classification for Security Operations Centers},
journal = {Journal of Information Security and Applications},
volume = {103},
pages = {104663},
year = {2026},
issn = {2214-2126},
doi = {10.1016/j.jisa.2026.104663},
url = {https://www.sciencedirect.com/science/article/pii/S2214212626002930},
author = {Risto Vaarandi and Alejandro Guerra-Manzanares}
}

@article{das2024effectiveness,
  author  = {Das, Shubhomoy and Islam, Md Rakibul and Jayakodi, Nitthilan Kannappan and Doppa, Janardhan Rao},
  title   = {Effectiveness of Tree-based Ensembles for Anomaly Discovery: Insights, Batch and Streaming Active Learning},
  journal = {Journal of Artificial Intelligence Research},
  year    = {2024},
  volume  = {80},
  pages   = {127--170},
  publisher = "AI Access Foundation",
  doi     = {10.1613/jair.1.14741}
}

@ARTICLE{zliobaite2014active,
  author={{\v{Z}}liobait{\.{e}}, Indr{\.{e}} and Bifet, Albert and Pfahringer, Bernhard and Holmes, Geoffrey},
  journal={IEEE Transactions on Neural Networks and Learning Systems}, 
  title={Active Learning With Drifting Streaming Data}, 
  year={2014},
  volume={25},
  number={1},
  pages={27-39},
  doi={10.1109/TNNLS.2012.2236570}}

@inproceedings{miller2016reviewer,
author = {Miller, Brad and Kantchelian, Alex and Tschantz, Michael Carl and Afroz, Sadia and Bachwani, Rekha and Faizullabhoy, Riyaz and Huang, Ling and Shankar, Vaishaal and Wu, Tony and Yiu, George and Joseph, Anthony D. and Tygar, J. D.},
title = {Reviewer Integration and Performance Measurement for Malware Detection},
year = {2016},
isbn = {9783319406664},
publisher = {Springer-Verlag},
address = {Berlin, Heidelberg},
url = {https://doi.org/10.1007/978-3-319-40667-1_7},
doi = {10.1007/978-3-319-40667-1_7},
booktitle = {Proceedings of the 13th International Conference on Detection of Intrusions and Malware, and Vulnerability Assessment - Volume 9721},
pages = {122–141},
numpages = {20},
location = {San Sebasti{\'a}n, Spain},
series = {DIMVA 2016}
}

@article{sethi2017md3,
title = {On the reliable detection of concept drift from streaming unlabeled data},
journal = {Expert Systems with Applications},
volume = {82},
pages = {77-99},
year = {2017},
issn = {0957-4174},
doi = {https://doi.org/10.1016/j.eswa.2017.04.008},
url = {https://www.sciencedirect.com/science/article/pii/S0957417417302439},
author = {Tegjyot Singh Sethi and Mehmed Kantardzic},
}

@inproceedings{drago2026online,
  author    = {Drago, Salvatore and
               Lorusso, Manuel and
               Ferraro, Pierluca and
               De Paola, Alessandra and
               Lo Re, Giuseppe},
  title     = {Online Framework for Intrusion Detection Systems with Explainable Concept Drift Detection and Adaptation Phases},
  booktitle = {Proceedings of the Joint National Conference on Cybersecurity (ITASEC \& SERICS 2026)},
  series    = {CEUR Workshop Proceedings},
  volume    = {4198},
  year      = {2026},
  publisher = {CEUR-WS.org},
  address   = {Cagliari, Italy},
  pages     = {1--15},
  numpages  = {15},
  url       = {https://ceur-ws.org}
}

@article{waschke2026ssl,
title = {Semi-supervised learning in Security Operations Centers: Performance, trade-offs and practical implications},
journal = {Computer Networks},
volume = {289},
pages = {112697},
year = {2026},
issn = {1389-1286},
doi = {https://doi.org/10.1016/j.comnet.2026.112697},
url = {https://www.sciencedirect.com/science/article/pii/S1389128626007097},
author = {Felix Waschke and Alejandro Guerra-Manzanares and Risto Vaarandi}
}

@article{waschke2026ml4soc,
author = {Waschke, Felix and Vaarandi, Risto and Guerra-Manzanares, Alejandro},
title = {ML4SOC: A Comprehensive Review on Machine Learning for Security Operations Centres},
year = {2026},
issue_date = {October 2026},
publisher = {Association for Computing Machinery},
address = {New York, NY, USA},
volume = {58},
number = {14},
issn = {0360-0300},
url = {https://doi.org/10.1145/3820494},
doi = {10.1145/3820494},
journal = {ACM Comput. Surv.},
month = jul,
articleno = {354},
numpages = {48},
}

\clearpage
\onecolumn
\appendix

\section{NAD-CD Upstream Signature Composition}
\label{app:signature_composition}
The NAD-CD evaluation stream contains 695 distinct upstream
\texttt{SignatureID} values, including 392 not present in FIT.
Figure~\ref{fig:appendix_signature_mix} shows the six most frequent
identifiers across ten consecutive chronological stream segments
containing approximately equal numbers of events, with all others
combined. Their relative frequencies change across replay order.
The identifiers are upstream detection metadata, not observed attack categories; these changes alone neither establish concept drift nor explain the Frozen screener's performance.

\setcounter{figure}{0}
\renewcommand{\thefigure}{A\arabic{figure}}
\begin{figure}[H]
  \centering
  \includegraphics[width=0.9\textwidth]{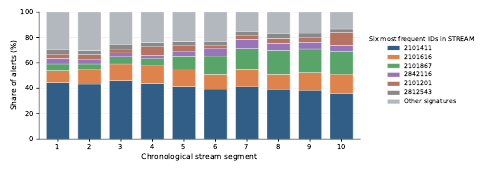}
  \caption{NAD-CD signature composition.}
  \label{fig:appendix_signature_mix}
  \Description{Stacked bars show the proportions of the six most frequent
  NAD-CD signature identifiers and all remaining identifiers in ten
  chronological stream segments containing approximately equal numbers
  of events, ordered by replay position.}
\end{figure}

\end{document}